\documentclass{article}

    \usepackage[final, nonatbib]{neurips_2023}

\usepackage[utf8]{inputenc} 
\usepackage[T1]{fontenc}    
\usepackage{hyperref}       
\usepackage{url}            
\usepackage{booktabs}       
\usepackage{amsfonts}       
\usepackage{nicefrac}       
\usepackage{microtype}      
\usepackage{xcolor}         
\usepackage{verbatim}       
\usepackage{graphicx}
\usepackage{multirow}
\usepackage{array}
\usepackage{amssymb}
\usepackage{tikz}
\usepackage{fontawesome5}
\usepackage{booktabs}
\usepackage{pifont}

\newcommand*\emptycirc[1][1ex]{\faCircle[regular]}
\newcommand*\halfcirc[1][1ex]{\faAdjust}
\newcommand*\fullcirc[1][1ex]{\faCheckCircle}

\title{\texttt{mir\_ref}: A Representation Evaluation Framework for Music Information Retrieval Tasks}

\author{
Christos Plachouras \quad\quad Pablo Alonso-Jim\'{e}nez \quad\quad Dmitry Bogdanov \\
Music Technology Group, Universitat Pompeu Fabra, Barcelona, Spain \\
\tt\small cplachouras@nyu.edu, pablo.alonso@upf.edu, dmitry.bogdanov@upf.edu
}

\begin{document}

\maketitle

\begin{abstract}
  Music Information Retrieval (MIR) research is increasingly leveraging representation learning to obtain more compact, powerful music audio representations for various downstream MIR tasks. However, current representation evaluation methods are fragmented due to discrepancies in audio and label preprocessing, downstream model and metric implementations, data availability, and computational resources, often leading to inconsistent and limited results. In this work, we introduce \texttt{mir\_ref},\footnote{\texttt{mir\_ref} is available at \url{https://github.com/chrispla/mir_ref}} an MIR Representation Evaluation Framework focused on seamless, transparent, local-first experiment orchestration to support representation development. It features implementations of a variety of components such as MIR datasets, tasks, embedding models, and tools for result analysis and visualization, while facilitating the implementation of custom components. To demonstrate its utility, we use it to conduct an extensive evaluation of several embedding models across various tasks and datasets, including evaluating their robustness to various audio perturbations and the ease of extracting relevant information from them.
\end{abstract}

\section{Evaluating Music Representations}

In the last decade, representation learning has attracted much interest in Music Information Retrieval (MIR), the field concerned with extracting, analyzing, and understanding information from music data. Time and frequency domain representations of audio are very information-dense, making it difficult and expensive to build end-to-end pipelines for solving MIR tasks. Additionally, their size and accompanying copyright restrictions make sharing, handling, and transferring music datasets challenging. Deep representations have shown promise as a generalized, compact, and efficient input feature relevant to many MIR tasks that circumvents these challenges. Many different music representation learning approaches have been undertaken, starting with the use of Deep Belief Networks in an unsupervised manner \cite{DBN_2009, DBN_2010, DBN_2011}. More popularly, classification approaches based on tags followed \cite{tagging_Dieleman_classification_2014, tagging_2015, tagging_2017, tagging_2019, Won_2020, tagging_Won_2020}, as well as some based on editorial metadata \cite{park_2017, lee_2019}. Correspondence has also been exploited, such as with tags \cite{multimodal_Favory_2020, multimodal_Favory_2021}, editorial metadata \cite{multimodal_Alonso_2022, COLA_2020}, playlists \cite{multimodal_Ferraro_2021, multimodal_Alonso_2023}, language \cite{multimodal_2021, multimodal_Manco_2022, mulan}, and video \cite{openl3}. A lot of interest has also fallen on self-supervised \cite{SSL_Spijkervet_2021, neuralfp, SSL_2020} and music-generation-based approaches \cite{SSL_Castellon_2021}.

Evaluation of music audio representations so far remains fragmented. This is in part attributable to challenges present in MIR like data unavailability and implementation complexity \cite{open-source_mir}, but, ultimately, there are no clear guidelines about how various components of a representation learning system should be implemented. Tools such as \texttt{mirdata} \cite{mirdata} and \texttt{mir\_eval} \cite{mir_eval} have encouraged consistency and transparency by standardizing dataset and metric implementations respectively. Practically, however, representation evaluation comes with many more components that need appropriate experimentation and transparent implementation. Table \ref{fig:implementations} demonstrates implementation choices by several works in the evaluation of their respective representation models in downstream classification tasks. Model sizes and optimization details vary significantly, and even preprocessing and prediction strategies are not consistent. In some cases, important implementation details are only present in the accompanying code, making them harder to find, or are missing from both the paper and code.

\begin{table}[!ht]
    \centering
    \caption{Downstream implementation details (selected). Downstream code and hyperparameter optimization (HPO) study open availability is indicated. Model output can be aggregated (aggr.) at the representation (repr.) or prediction (pred.) level. \texttt{\textbf{?}} indicates implementation detail is missing. }
    \label{fig:implementations}
    \begin{tabular}{lccccccc}
    \hline
        ~ & ~ & \multicolumn{2}{c}{\centering model} & \multicolumn{3}{c}{\centering optimization} & output \\ 
         \noalign{\vskip -2pt} 
        \cmidrule(lr){3-4} \cmidrule(lr){5-7} \cmidrule(lr){8-8}
        \noalign{\vskip -1.4pt}
        ~ & code & type & layer(s) & HPO & initial $lr$ & $wd$ & aggr. \\ \hline 
        \noalign{\smallskip}
        EffNet-Discogs & ~ & MLP & 512 & ~ & $1e^{-3}$ & $1e^{-5}$ & pred. \\ 
        MusiCNN & \checkmark & SVM & NA & ~ & NA & NA & pred. \\ 
        OpenL3 & ~ & MLP & 512-128 & \checkmark & $1e^{\{-5,..,-3\}}$ & $1e^{\{-5,..,-3\}}$ & pred. \\ 
        NeuralFP  & ~ & LC & NA & ~ & ? & ? & ? \\ 
        CLMR & \checkmark & LC & NA & ~ & $3e^{-4}$ & $1e^{-6}$ & repr. \\ 
        MERT  & \checkmark & MLP & 512 & \checkmark & $1e^{\{-4,..,-2\}}$ & ? & repr. \\
        COALA & \checkmark & MLP & 256 & ~ & $1e^{-3}$ & $1e^{-4}$ & repr. \\
        JukeMIR & \checkmark & LC/MLP & NA/512 & \checkmark & $1e^{\{-5,..,-3\}}$ & $1e^{\{-3,..,0\}}$ & repr. \\
        MuLaP & \checkmark & MLP & 512 & ~ & $1e^{-3}$ & $1e^{-2}$ & pred. \\
        \hline
    \end{tabular}
    
\end{table}


Additionally, setting up evaluation experiments is tedious. Getting access to and handling several large datasets is challenging, and ensuring adequate experimentation on downstream pipeline components and parameters is time-consuming and, often, computationally infeasible. As a result, a limited set of datasets and tasks is usually investigated in the original representation model works (see Table \ref{fig:datasets}).

\begin{table}[!ht]
    \centering
    \caption{Music tasks and datasets used for evaluation in the original representation model works.}
    \label{fig:datasets}
    \begin{tabular}{lccccccccccccccc}
        \rotatebox{90}{~} & \rotatebox{90}{MTG J. genre} & \rotatebox{90}{MTG J. instr.} & \rotatebox{90}{MTG J. mood} & \rotatebox{90}{MTG J. top50} & \rotatebox{90}{GTZAN genre} & \rotatebox{90}{MTAT tagging} & \rotatebox{90}{MSD tagging} & \rotatebox{90}{FMA genre} & \rotatebox{90}{FMA identity} & \rotatebox{90}{EMO emotion} & \rotatebox{90}{GS key} & \rotatebox{90}{NSynth instr.} & \rotatebox{90}{Nsynth pitch} & \rotatebox{90}{Vocalset singer} & \rotatebox{90}{Vocalset tech.} \\ \hline
        \noalign{\smallskip}
        EffNet-Discogs & \checkmark & \checkmark & \checkmark & \checkmark & ~ & \checkmark & ~ & \checkmark & ~ & ~ & ~ & ~ & ~ & ~ & ~ \\ 
        MusiCNN & ~ & ~ & ~ & ~ & \checkmark & ~ & ~ & ~ & ~ & ~ & ~ & ~ & ~ & ~ & ~ \\ 
        NeuralFP & ~ & ~ & ~ & ~ & \checkmark & ~ & ~ & ~ & \checkmark & ~ & ~ & ~ & ~ & ~ & ~ \\ 
        CLMR & ~ & ~ & ~ & ~ & ~ & \checkmark & \checkmark & ~ & ~ & ~ & ~ & ~ & ~ & ~ & ~ \\ 
        MERT & \checkmark & \checkmark & \checkmark & \checkmark & \checkmark & \checkmark & ~ & ~ & ~ & \checkmark & \checkmark & \checkmark & \checkmark & \checkmark & \checkmark \\ 
        COALA & ~ & ~ & ~ & ~ & \checkmark & ~ & ~ & ~ & ~ & ~ & ~ & \checkmark & ~ & ~ & ~ \\ 
        JukeMIR & ~ & ~ & ~ & ~ & \checkmark & \checkmark & ~ & ~ & ~ & \checkmark & \checkmark & ~ & ~ & ~ & ~ \\ 
        MuLaP & \checkmark & \checkmark & \checkmark & \checkmark & ~ & \checkmark & ~ & \checkmark & ~ & \checkmark & ~ & \checkmark & ~ & ~ & ~ \\ 
    \end{tabular}
\end{table}

Benchmarks-challenges, historically popular in MIR \cite{mirex_10_years}, have helped alleviate the problem of evaluation result comparability. The Holistic Evaluation of Audio Representations (HEAR) benchmark \cite{HEAR} contains several music-related tasks that can be used to compute results from a representation and submit them to the leaderboard. Similarly, the Holistic Audio Representation Evaluation Suite (HARES) \cite{HARES} and the Evaluation Package for Audio Representations \cite{EVAR_github} contain a few music-related tasks with a fixed downstream pipeline for evaluation. Recently, a representation evaluation benchmark specifically aimed at MIR tasks was released, called the Music Audio Representation Benchmark for universaL Evaluation (MARBLE) \cite{marble}. MARBLE implements a wide range of MIR tasks and datasets, and is primarily submission-based. A fixed downstream setup is also enforced, with a one-layer 512-unit MLP for all tasks apart from source separation.

While these efforts contribute towards consistent representation evaluation, there are some inherent limitations to such benchmarks. Typically, they evaluate representation learning systems within a constrained downstream environment. This rigidity means that potential optimizations, such as modifying the downstream model's structure, are overlooked. Moreover, while benchmarks offer a streamlined approach to evaluation, they might not fully capture a system's real-world performance. Crucial aspects like a system's adaptability to audio deformations, performance with novel data, computational constraints, and other nuanced factors might be left unexplored. Notably, the submission-first nature of many benchmarks makes them unsuitable for use as a local aid for representation development. Therefore, while benchmarks serve as valuable standardized tools for evaluation, they cannot entirely supplant the need for more comprehensive assessments of representation learning systems.

\section{Overview of \texttt{mir\_ref}}

\texttt{mir\_ref} is a Python framework designed for holistic and transparent evaluation of music representations. It features a diverse array of MIR tasks, datasets, embedding models, metrics, and other essential components and follows a configuration-based approach that allows experiment design and conduct without code and data handling. At the same time, it is designed to facilitate custom component integration with minimal interfacing with the rest of the framework, encouraging contributions and facilitating the incorporation of proprietary datasets and models. A core focus of the framework is to make most of the parameters and implementations along the transfer learning pipeline transparent and available for experimentation. This is because many questions remain open regarding the performance implications of other components such as the embedding extraction window frequency, the embedding preprocessing used, the downstream model structure, the optimizer configuration, the embedding or prediction aggregation, and others. As will be demonstrated in section \ref{sec:example_experiments}, these parameters can withhold a lot of performance from the representation. 

\begin{figure}[!ht]
    \centering
    \includegraphics[width=1\linewidth]{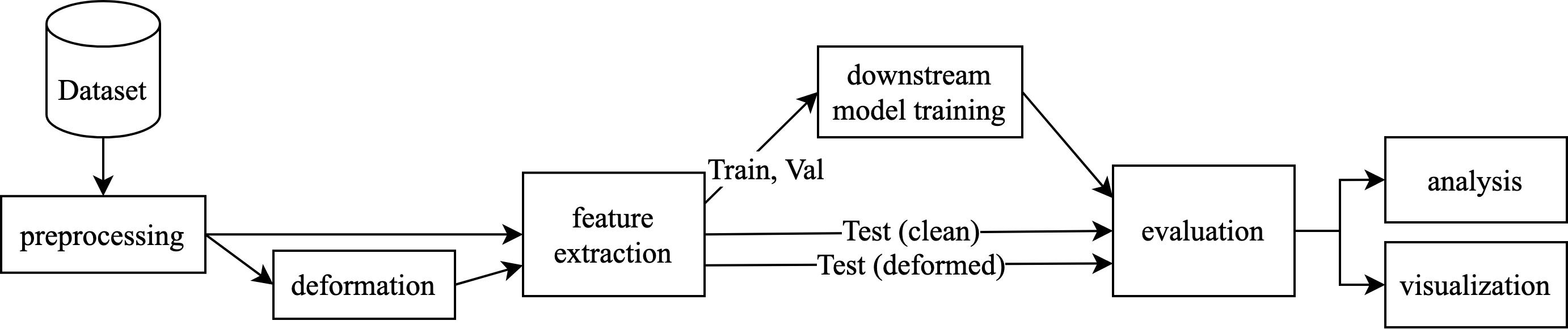}
    \caption{Flowchart of the main experiment components of \texttt{mir\_ref}}
    \label{fig:flowchart}
\end{figure}

\texttt{mir\_ref} experiments are comprised of several main components, as seen in Fig. \ref{fig:flowchart}, which will be described at a high level due to space constraints. We always recommend that the latest documentation and guides are consulted.

\paragraph{Datasets} Datasets are the core component of every run. During an experiment run, dataset setup including downloading, preprocessing, loading, handling, and interfacing with task components happens automatically, without any user intervention. We leverage \texttt{mirdata} \cite{mirdata_zenodo} for implementing many of its available datasets, ensuring consistent and reliable dataset handling. Due to the absence of some datasets in \texttt{mirdata} and to provide the ability to easily implement custom datasets, we wrap relevant functionality in a \texttt{mir\_ref} dataset class. However, we encourage users to contribute dataset handlers directly to \texttt{mirdata}, and interface with the \texttt{mir\_ref} dataset class only for quick, custom dataset implementations.

\paragraph{Deformations} Deformations are a core component of \texttt{mir\_ref} as robustness evaluation is critical in assessing real-world system performance. They are computed using \texttt{audiomentations} \cite{audiomentations}, immediately making dozens of relevant deformations available. By default, representations of deformed audio are used exclusively in the evaluation process and not while training the downstream models to assess the robustness of the representation itself.

\paragraph{Feature extraction} Several representation models are provided in \texttt{mir\_ref} for use as feature extractors, some of which are implemented with Essentia models \cite{essentia_models}. It is worth noting that non-learned representation can also be utilized, and some are already implemented using Essentia \cite{essentia} primarily for use as baselines.

\paragraph{Downstream models} A utility is provided to construct models based on the parameters in the configuration file. The utility can adjust models accordingly depending on the embedding size, such as, for example, when embedding dimensionality is used, or when different embeddings in an experiment have different dimensions. As is the case with all other components, the user can easily implement and use their own downstream model. Tensorboard \cite{tensorflow} is currently used to visualize training logs.

\paragraph{Evaluation, analysis, and visualization} For each task, a set of default evaluation metrics is available, implemented with \texttt{mir\_eval} \cite{mir_eval} when applicable. Because a huge amount of information can be produced by some extensive experiments, a tool to create tables based on the results and a selected preset is provided, which we plan to make interactive in the next versions of the framework. An interactive confusion matrix that provides audio and spectrogram references for misclassified data is also provided to help better understand system behavior. We plan to develop more tools leveraging the same results format to aid representation understanding and explainability.

\section{Example experiments}\label{sec:example_experiments}
We conducted a demonstrative evaluation using 7 representation models, 6 datasets and tasks, 4 audio deformations, and 5 downstream model configurations. Due to space constraints, we reference a few notable results, and we make the full results, an analysis of them, and the \texttt{mir\_ref} configuration file to reproduce them available through the framework's code repository.

From our experiments, we found that these representations generally 
struggle with audio deformations like white noise and gain reduction, 
though they fare better with intense MP3 compression. This resilience 
varies between tasks, with larger downstream models often compensating for some 
loss in performance. For instance, in instrument recognition using the 
TinySOL dataset \cite{orchideasol}, a dataset of monophonic, single-note strokes from different instruments, white noise at 15 dB SNR sees a significant drop in performance for CLMR and OpenL3. At an SNR of 0 dB, all models falter. 
Meanwhile, in key detection with the Beatport EDM dataset, while OpenL3 
and VGGish remain largely stable, NeuralFP and CLMR, which are trained 
to be noise-invariant, notably struggle under extreme noise.

Another revealing outcome of these experiments is how much impact the downstream model choice can have. We used an SLP, as well as 4 MLPs, 2 of which had a fixed size ({128} and {256, 128}) and 2 which adapted their size based on the representation's shape. For singer identification in VocalSet \cite{vocalset}, a dataset of monophonic singing recordings, most models have a significant performance difference between the linear classifier and larger models, with cases like NeuralFP doubling their F1 score, although others like MERT exhibit no notable performance differences. This points to the fact that singer identity information might not be linearly separable in some of these representations, and generally also suggests that performance on different downstream models might be an indicator of how easy relevant information is to extract from a representation.

Lastly, we implemented a pitch class classification task on TinySOL to gauge whether pitch information is encoded in the representations. We observe that almost all representations are unsuitable for predicting pitch class, even in this simple scenario of monophonic, single-note instrument strokes. The only two representations able to discern pitch were from MERT, possibly aided by the music teacher used during training, and NeuralFP, suggesting that it could be relying heavily on melody as a proxy for identification. As interest in creating more generalizable representations grows, it is interesting to investigate how input representations and training strategies affect the music information that is possibly encoded in a learned representation.

\section{Discussion}
These results serve as encouragement to start thinking beyond benchmarks when it comes to music representation evaluation. There remains much to investigate and understand about how to optimize, comprehend, and interpret learned representations. At the same time, designing and conducting extensive, holistic evaluation experiments needs to be done in a transparent and reproducible manner. We believe \texttt{mir\_ref} is a step in that direction. As we keep developing this framework, we want to engage with the MIR community to better understand use cases as well as individual workflows. At the same time, we hope the modular, custom component interface provides a low barrier to entry for contributions while supporting existing standardization efforts in MIR.


\begin{ack}
This work has been supported by the Musical AI project - PID2019-111403GB-I00/AEI/10.13039/\ 501100011033, funded by the Spanish Ministerio de Ciencia e Innovaci{\'o}n and the Agencia Estatal de Investigaci{\'o}n.

\end{ack}

\bibliographystyle{plain} 
{\small \bibliography{refs}} 

\end{document}